\documentclass[aps]{revtex4}

\usepackage{graphicx}
\usepackage{dcolumn}
\usepackage{bm}

\usepackage[english]{babel}
\begin{document}

\title{\large\bf A Precision Measurement of the Mass of the Top Quark}

%
\author{D\O\ Collaboration$^{*}$}

\affiliation{$^{*}$A list of authors and their affiliations appears at the end of the paper}
                                                                 
\rm\footnotesize

\maketitle

\rm\normalsize


{\bf 
The Standard Model (SM) of particle physics contains about 
two dozen parameters~-- such as particle masses~-- whose 
origins are still unknown and cannot be predicted, but whose values 
are constrained through their interactions. In
particular, the masses of the top ($\bm t$) quark ($\bm{M_t}$) and $\bm W$
boson ($\bm{M_W}$) \cite{pdg3} constrain the mass of the long-hypothesized,
but thus far not observed, Higgs boson.
A precise measurement of the top-quark mass can therefore point to where
to look for the Higgs, and indeed whether the hypothesis of a SM Higgs 
is consistent with experimental data. Since top quarks are produced 
in pairs and decay in only $\bm{\approx 10^{-24}}$~s into various final states, reconstructing their mass
from their decay products is very challenging. Here we report a technique that extracts far more information from each top-quark event and yields a greatly improved precision on the top mass of $\bm{\pm 5.3}$~GeV/$\bm{c^2}$, compared to
previous measurements~\cite{cdfljets}. When our new result is combined with 
our published measurement in a complementary decay mode~\cite{dilep11} and with 
the only other measurements available~\cite{cdfljets}, the new world average for $\bm M_t$ becomes $\bm{178.0 \pm 4.3}$ GeV/$\bm{c^2}$~\cite{wa}.
As a result, the most likely Higgs mass increases from the experimentally excluded~\cite{lepdirect} value of 96~GeV/$\bm{c^2}$~\cite{lepEWWG4} to 117~GeV/$\bm{c^2}$, which is beyond current experimental sensitivity. The upper limit on the Higgs mass at 95\% confidence level is raised from 219~GeV/$\bm{c^2}$ to 251~GeV/$\bm{c^2}$.
}
\medskip

The discovery of the top quark in 1995 served as one of the major
confirmations  of the validity of the SM \cite{cdfPRD1,d0PRD2}. 
Of its many parameters, the mass of the top quark, in particular, 
reflects some of the most crucial aspects of the Model. This is 
because, in principle, the top quark is point-like and should be massless; yet, through its interactions with the hypothesized Higgs field, the physical mass of the top quark appears to be about the 
mass of a gold nucleus. Because it is so heavy, the top quark (along with the $W$ boson) provides an unusually sensitive tool for investigating the Higgs field. 
$M_W$ is known to a precision of $0.05\%$, while the uncertainty 
on $M_t$ is at the 3$\%$ level \cite{pdg3}. Improvements in both 
measurements are required to restrict further the allowed range 
of mass for the Higgs; however, given the large uncertainty in the top mass, an improvement in its precision is particularly important. 
As has been pointed out recently, e.g., in Refs.~\cite{Higgs1,Higgs2}, a potential problem for the 
SM is that, based on the presently accepted mass of the top quark,
the most likely value of the Higgs mass~\cite{lepEWWG4} lies in a range that has already been excluded by experiment \cite{lepdirect}. 
Precise knowledge of the Higgs mass is crucial 
for our understanding of the SM and any possible new physics beyond it.
For example, in a large class of supersymmetric models (theoretically preferred
solutions to the deficiencies of the SM), the Higgs mass 
has to be less than $\approx 135$ GeV/$c^2$. While, unlike the SM, supersymmetry 
predicts more than one Higgs boson, the properties of the lightest
one are expected to be essentially the same as those for the SM Higgs boson. 
Thus, if the SM-like Higgs is heavier than $\approx 135$~GeV/$c^2$, it would disfavor a large class of supersymmetric models. In addition,
the current limits on supersymmetric particles from LEP~\cite{LEPSUSY} are extremely sensitive to the mass of the top quark. In fact, for a top-quark mass greater than 179~GeV/$c^2$, the bounds on one of the major supersymmetry parameters, $\tan\beta$, which relates the properties of the SM-like Higgs boson and its heavier partners, would disappear completely~\cite{degrassi}. 
Hence, in addition to the impact on searches for the Higgs boson, other important consequences call for improved precision on the top mass, and this goal is the main subject of this paper.

\begin{figure}[htbp]
\begin{center}
\includegraphics[width=.7\textwidth]{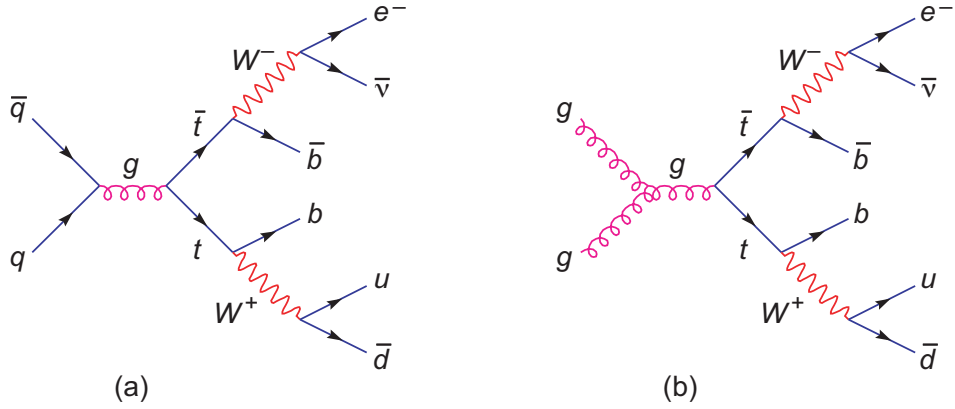}
\end{center}
\vskip -0.5cm
\caption[]{Feynman diagrams for $t\bar t$ production in $p\bar p$ collisions, with subsequent decays into an electron, neutrino, and quarks. Diagram (a) (quark-antiquark production) is dominant, but diagram (b) (gluon fusion) contributes $\approx 10\%$ to the cross section. This particular final state ($e\bar\nu u \bar d b\bar b$) is one of the channels used in the analysis.}
\label{fig:production}
\end{figure}

\begin{figure}[htbp]
\begin{center}
\includegraphics[width=.6\textwidth]{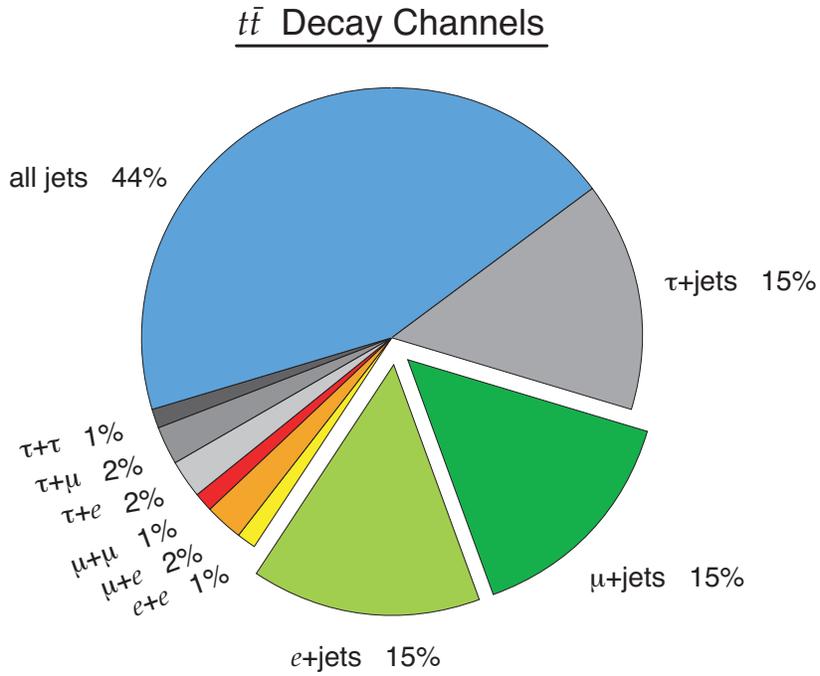}
\end{center}
\vskip -0.5cm
\caption[]{Relative importance of various $t\bar t$ decay modes. The ``lepton $+$ jets'' channel used in this analysis corresponds to the two offset slices of the pie-chart and amounts to 30\% of all the $t\bar t$ decays.}
\label{fig:decay}
\end{figure}

The D\O\ experiment at the 
Fermilab Tevatron has studied a sample of $t \bar t$ events produced in proton-antiproton ($p\bar p$)
interactions~\cite{massPRD5}. The total energy of 1.8 TeV released in a head-on collision of a
900 GeV $p$ and 900 GeV $\bar p$ is almost as large as the rest energy of ten gold nuclei. Each top (antitop) quark decays almost immediately into a bottom $b$ ($\bar b$) 
quark and a $W^+$ ($W^-$) boson, and we have reexamined those events 
in which one of the $W$ bosons decays into a charged lepton (electron or 
muon) and a neutrino, and the other $W$ into a quark and an antiquark (see Figure~\ref{fig:production}). 
These events and their
selection criteria are identical to those used to extract the mass 
of the top quark in our previous publication, and correspond to an 
integrated luminosity of 125 events/pb. (That is, given the production cross
section of the $t\bar t$ in $p\bar p$ collisions at 1.8 TeV of 5.7 pb, as measured by D\O\ \cite{topcs},
these data correspond to approximately 700 produced $t\bar t$ pairs, a fraction of which is fully detected in various possible decay modes. Approximately 30\% of these correspond to the ``lepton $+$ jets'' topology categorized in Figure~\ref{fig:decay}, where ``jet'' refers to products of the fragmentation of a quark into a collimated group of particles that are emitted along the quark's original direction.) The main background processes correspond to multijet production (20\%), where one of the jets is reconstructed incorrectly as a lepton, and the $W +$ jets production with leptonic $W$ decays (80\%), which has the same topology as the $t\bar t$ signal.

The previous D\O\ top-quark mass measurement in this lepton $+$ jets channel is $M_t = 173.3 \pm 5.6~\mbox{(stat)} \pm 5.5~\mbox{(syst)}$~GeV/$c^2$, and is based on 91 candidate events. Information pertaining to the detector and to the older analysis can be found in Refs. \cite{d0NIM6} and \cite{massPRD5}, respectively.

The new method of the top-quark mass measurement is similar to one suggested previously \cite{dgk10,dgk11} for $t \bar t$ dilepton decay
channels (where both $W$ bosons decay leptonically), and used in previous mass analyses of dilepton events
\cite{dilep11}, and akin to an approach suggested for the measurement of 
the mass of the $W$ boson at LEP \cite{berends11.5,delphi,juste}.  
The critical differences from previous analyses in the lepton $+$ jets decay channel lie in: (i) the assignment of more 
weight to events that are well measured or more likely to correspond to $t \bar t$ signal, and (ii) the handling of the combinations of final-state objects
(lepton, jets, and imbalance in transverse momentum, the latter being a signature for an undetected neutrino) and their identification with
top-quark decay products in an event (e.g., from ambiguity in choosing jets that correspond to $b$ or $\bar b$ quarks from the decays of the $t$ and $\bar t$ quarks). Also, since leading-order matrix elements were used to calculate the event weights, only events with exactly four jets are kept in this analysis, resulting in a candidate sample of 71 events. Although we are left with fewer events, the new method for extracting the mass of the top quark provides substantial improvement in both statistical and systematic uncertainties.

We calculate as a function of top mass the differential probability that the measured variables in any event correspond to signal. The maximum in the product of these individual event probabilities provides the best estimate of the mass of the top quark in the data sample.
The impact of biases from imperfections in the detector and event reconstruction algorithms is taken into account in two ways. Geometric acceptance, trigger efficiencies, event selection, etc., enter through a multiplicative acceptance function that is independent of $M_t$. Because the angular directions of all the objects in the event, as well as the electron momentum, are measured with high precision, their measured values are used directly in the calculation of the probability that any event corresponds to $t \bar t$ or background production. The known momentum resolution is used to account for uncertainties in measurements of jet energies and muon momenta. 

As in the previous analysis \cite{massPRD5}, momentum conservation in $\gamma+$jet events is used to check that the energies of jets in the experiment agree with Monte Carlo (MC) simulation. This calibration has an uncertainty $\delta E= (0.025 E + 0.5$ GeV). Consequently, all jet energies in our sample are rescaled by $\pm\delta E$, the analysis redone, and half of the difference in the two rescaled results for $M_t$ ($\delta M_t = 3.3$ GeV/$c^2$) is taken as a systematic uncertainty from this source. All other contributions 
to systematic uncertainty: Monte Carlo modeling of signal ($\delta M_t = 1.1$ GeV/$c^2$) and background ($\delta M_t = 1.0$ GeV/$c^2$), effect of calorimeter noise and event pile-up ($\delta M_t = 1.3$ GeV/$c^2$), and other corrections from top-quark mass extraction ($\delta M_t = 0.6$ GeV/$c^2$) are much smaller, and discussed in detail in Refs.~\cite{thesis1,thesis2}. It should be noted that the new mass measurement method provides a significant ($\approx 40\%$, from $\pm 5.5$ to $\pm 3.9$ GeV/$c^2$) reduction in systematic uncertainty, which is ultimately dominated by the measurement of jet energies. For details on the new analysis, see the Methods section. 

The final result is $M_t=180.1\pm 3.6~\mbox{(stat)} \pm 3.9$ (syst) GeV/$c^2$. The improvement in statistical uncertainty over our previous measurement 
is equivalent to collecting a factor of 2.4 as much data. Combining the statistical and systematic uncertainties in quadrature, we obtain $M_t=180.1 \pm 5.3$ GeV/$c^2$, which is consistent with our previous measurement in the same channel (at $\approx 1.4$ standard deviations), and has a precision comparable to all previous top-quark mass measurements combined \cite{pdg3}.

\begin{figure}[htbp]
\begin{center}
\includegraphics[width=.65\textwidth]{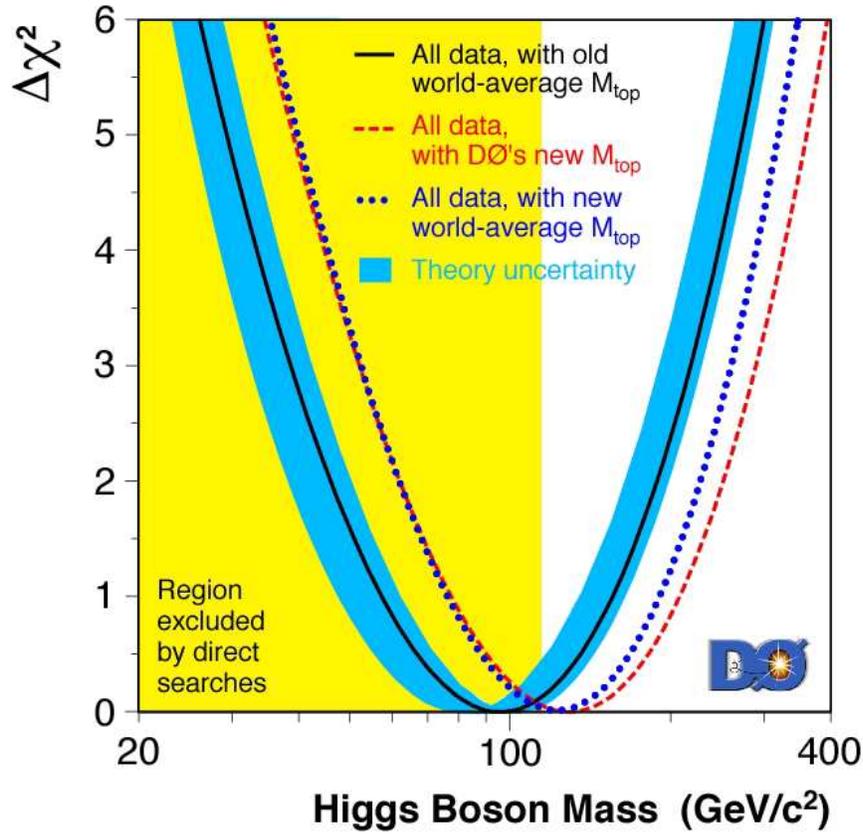}
\end{center}
\vskip -0.5cm
\caption[]{Current experimental constraints on the mass of the Higgs boson.
The $\chi^2$ for a global fit to electroweak data 
using the procedure of Ref.~\protect\cite{lepEWWG4}, is shown as a 
function of the Higgs mass. The solid line corresponds to 
the result for the previous world average for the top-quark mass of $174.3 \pm 5.1$ GeV/$c^2$, with the blue band indicating the impact of theoretical uncertainty. 
The dotted line shows the result for the new world-averaged $M_t$ of $178.0 \pm 4.3$~GeV/$c^2$, while the dashed line corresponds to using just the new D\O\ average  of $179.0 \pm 5.1$~GeV/$c^2$. 
The yellow shaded area on the left indicates
the region of Higgs masses excluded by experiment ($> 114.4$~GeV/$c^2$ at 
the 95\% confidence level~\protect\cite{lepdirect}). The improved top 
mass measurement shifts the most likely value of the Higgs mass above the 
experimentally excluded range.} 
\label{fig:blueband}
\end{figure}

The new measurement can be combined with that obtained for the 
dilepton sample that was also collected at D\O\ during Run I \cite{dilep11}
($M_t = 168.4 \pm 12.3~\mbox{(stat)} \pm 3.6~\mbox{(syst)}$~GeV/$c^2$), to yield the new D\O\ average for the mass of the top quark:
\begin{equation}
M_t=179.0 \pm 5.1\mbox{ GeV/$c^2$ (D\O)}.
  \nonumber
\end{equation}
Combining this with measurements from the CDF experiment~\cite{cdfljets}, provides a new ``world average'' (based on all measurements available) for the top-quark mass~\cite{wa}:
\begin{equation}
M_t=178.0 \pm 4.3\mbox{ GeV/$c^2$ (All available data)},
\end{equation}
dominated by our new measurement.
This new world average shifts the best-fit value of the expected Higgs mass from 
96~GeV/$c^2$ to 117~GeV/$c^2$ (see Figure~\ref{fig:blueband}), which is now outside of the experimentally-excluded region, yet accessible in the current run of the Tevatron and at future runs at the Large Hadron Collider (LHC), currently under construction at CERN.  (The upper limit on the Higgs mass at 95\% confidence level changes from 219~GeV/$c^2$ to 251~GeV/$c^2$.)
Figure~\ref{fig:blueband} shows the effect of using only the new D\O\ top mass for fits to the Higgs mass, and indicates a best value of 123~GeV/$c^2$ and the upper
limit of 277~GeV/$c^2$ at 95\% confidence level. It should be noted that the horizontal scale in Figure~\ref{fig:blueband} is logarithmic, and the limits on the Higgs boson mass are therefore asymmetric.

The new method is already being applied to data being collected by the CDF and D\O\ experiments at the new run of the Fermilab Tevatron and should provide even higher precision on the determination of the top-quark mass, equivalent to more than a doubling of the data sample, relative to using the conventional method. An ultimate precision of at $\approx 2$~GeV/$c^2$ on the top-quark mass is expected to be reached in several years of Tevatron operation. Further improvement may eventually come from the LHC.

\section*{\bf Methods} 

The probability density as a function of $M_t$ can be written 
as a convolution of the calculable cross section and any effects from 
measurement resolution:
\begin{equation}
P(x,M_t) = \frac{1}{\sigma(M_t)} \int {\textup d} \sigma(y,M_t) 
{\textup d} q_1 {\textup d} q_2 f(q_1) f(q_2) W(y,x) 
\label{G_P(x)}
\end{equation}
where $W(y,x)$, our general transfer function, is the normalized probability 
for the measured set of variables $x$ to arise from a set of nascent 
(partonic) variables $y$, $d\sigma(y,M_t)$ is the partonic theoretical
differential cross section, $f(q)$ are parton distribution 
functions that reflect the probability of finding any specific 
interacting quark (antiquark) with momentum $q$ within the proton (antiproton), and $\sigma(M_t)$ is the total cross section for producing $t\overline{t}$.
The integral in Eq. (\ref{G_P(x)}) sums over all possible parton states 
leading to what is observed in the detector.  

The acceptance of the detector is given in terms of a function $A(x)$ that relates the probability $P_m(x,M_t)$ of measuring the observed variables $x$ to their production probability $P(x,M_t)$: $P_m(x,M_t) = A(x) P(x,M_t)$. Effects from energy resolution, etc., are taken into account in the transfer function $W(y,x)$. The integrations in Eq. (\ref{G_P(x)}) over the eleven well-measured variables (three components of charged-lepton momentum and eight jet angles) and the four equations of energy-momentum conservation, leave five integrals that must be performed to obtain the probability that any event represents $t \overline{t}$ (or background) production for some specified value of top mass $M_t$.

The probability for a $t \bar t$ interpretation can be written as:
$$
P_{t \bar t}=\frac{1}{12 \sigma_{t \bar t}} \int d^5 \Omega
 \sum_{\text{perm.},\nu} 
 |{\cal M}_{t \bar t}|^2 \frac{f(q_1)f(q_2)}{|q_1||q_2|} \Phi_6 
W_{\text{jets}}(E_{\text{part}},E_{\text{jet}}),
$$
where $\Omega$ represents a set of five integration variables, 
${\cal M}_{t \bar t}$ is the leading-order matrix element for $t\bar t$ production \cite{sparke1,sparke2}, $f(q_1)$ and $f(q_2)$ are the CTEQ4M parton distribution functions for the incident quarks \cite{cteq}, $\Phi_6$ is the phase-space factor for the 6-object final state, and the sum is over all 12
permutations of the jets and all possible neutrino solutions.
$W_{\text{jets}}(E_{\text{part}},E_{\text{jet}})$ corresponds
to a function that maps parton-level energies $E_{\text{part}}$
to energies measured in the detector $E_{\text{jet}}$, and is
based on MC studies. A similar expression, using a
matrix element for $W +$ jets production (the dominant background source) that is independent of $M_t$, is used to calculate the probability for a background interpretation, $P_{\text{bkg}}$.

Studies of samples of {\sc HERWIG}~\cite{HERWIG} MC events indicate that the new method is capable of providing 
almost a factor of two reduction in the statistical uncertainty 
on the extracted $M_t$. These studies also reveal that there is 
a systematic shift in the extracted $M_t$ that depends on the 
amount of background there is in the data. To minimize this effect, a 
selection is introduced based on the probability that an event 
represents background. The specific value of 
the $P_{\text{bkg}}$ cutoff is based 
on MC studies carried out before applying the method to data, 
and, for a top mass of 175 GeV/$c^2$, retains 71$\%$ of the signal 
and 30$\%$ of the background. A total of 22 data events out of our 71 candidates pass this selection.

The final likelihood as a function of $M_t$ is written as:
$$
\ln L(M_t) = \sum_{i=1}^N \ln[c_1 P_{t \bar t}(x_i, M_t) +
c_2 P_{\text{bkg}}(x_i)]
          - N \int A(x) \left[ c_1 P_{t \bar t}(x, M_t) +
            c_2 P_{\text{bkg}}(x)\right] {\textup d}x,
$$
The integration is performed using MC methods. The best 
value of $M_t$ (when $L$ is at its maximum $L_{\mbox{max}}$) represents the most likely mass of top in the 
final $N$-event sample, and the parameters $c_i$ reflect the amounts
of signal and background. MC studies show that
there is a downward shift of 0.5 GeV/$c^2$ in the extracted mass, and this
correction is applied to the result. Reasonable changes in the
cutoff on  $P_{\text{bkg}}$ do not have significant impact on $M_t$.

\begin{figure}[htbp]
\begin{center}
\includegraphics[width=.65\textwidth]{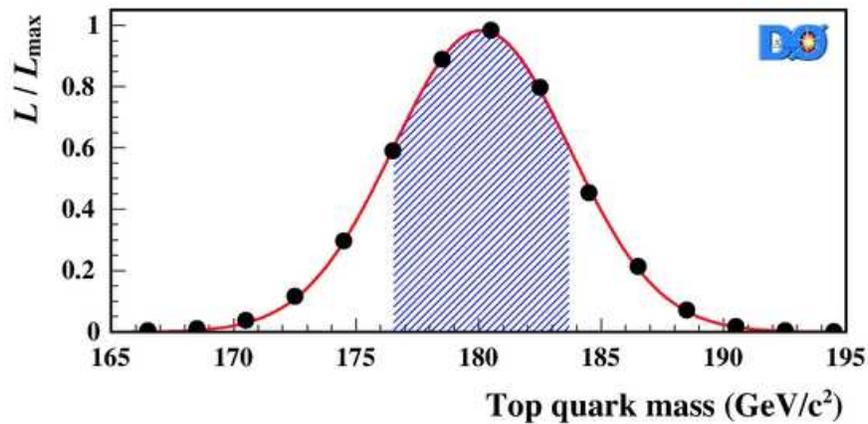}
\end{center}
\vskip -0.5cm
\caption[]{Determination of the top-quark mass using the maximum likelihood method.
The points represent the likelihood of the fit used to extract the top mass, divided by its maximum value, as a function of the mass of the top quark (after a correction for a $-0.5$~GeV/$c^2$ mass bias, see text). The solid line shows a Gaussian fit to the points. The maximum likelihood corresponds to a mass of 180.1~GeV/$c^2$, which is the new D\O\ measurement of the top mass in the lepton $+$ jets channel. The hatched band corresponds to the range of $\pm 1$ standard deviation, and indicates the $\pm 3.6$~GeV/$c^2$ statistical uncertainty of the fit.}
\label{fig:lik_topW}
\end{figure}

Figure~\ref{fig:lik_topW} shows the value of $L(M_t)/L_{\text{max}}$ as a 
function of $M_t$ for the 22 events that pass all selection criteria, 
after correction for the $0.5$ GeV/$c^2$ bias in mass. The likelihood is
maximized with respect to the parameters $c_i$ at each mass 
point. The Gaussian fit in the figure yields $M_t = 180.1$ GeV/$c^2$,
with a statistical uncertainty of $\delta M_t = 3.6$ GeV/$c^2$. The systematic uncertainty, dominated by the measurement of jet energies, as discussed above, amounts to $\delta M_t = 3.9$ GeV/$c^2$. When added in quadrature to the statistical uncertainty from the fit, it yields the overall uncertainty on the new top-quark mass measurement of $\pm 5.3$~GeV/$c^2$.

\medskip
We wish to note the great number of contributions made by the late Harry Melanson
to the D\O\ experiment, through his steady and inspirational leadership of the physics, reconstruction and algorithm efforts.

\medskip

Correspondence and requests for material should be addressed to J.~Estrada (estrada@fnal.gov).

\medskip
%
We are grateful to our colleagues Arnulf Quadt and Martijn Mulders for careful reading of the manuscript and detailed comments. 
We thank the staffs at Fermilab and collaborating institutions, 
and acknowledge support from the 
Department of Energy and National Science Foundation (USA),  
Commissariat  \` a L'Energie Atomique and 
CNRS/Institut National de Physique Nucl\'eaire et 
de Physique des Particules (France), 
Ministry for Science and Technology and Ministry for Atomic 
   Energy (Russia),
CAPES, CNPq and FAPERJ (Brazil),
Departments of Atomic Energy and Science and Education (India),
Colciencias (Colombia),
CONACyT (Mexico),
Ministry of Education and KOSEF (Korea),
CONICET and UBACyT (Argentina),
The Foundation for Fundamental Research on Matter (The Netherlands),
PPARC (United Kingdom),
Ministry of Education (Czech Republic),
A.P.~Sloan Foundation,
and the Research Corporation.
%



\bigskip

\noindent
V.M.~Abazov$^{22}$,                                                           
B.~Abbott$^{55}$,                                                             
A.~Abdesselam$^{11}$,                                                         
M.~Abolins$^{48}$,                                                            
V.~Abramov$^{25}$,                                                            
B.S.~Acharya$^{17}$,                                                          
D.L.~Adams$^{53}$,                                                            
M.~Adams$^{35}$,                                                              
S.N.~Ahmed$^{21}$,                                                            
G.D.~Alexeev$^{22}$
% ...
\\                                                                            
\medskip

\noindent                                                                     
1, Universidad de Buenos Aires, Departamento de F{\'\i}sica, FCEN, 
Pabell\'{o}n 1, Ciudad Universitaria, 1428 Buenos Aires, Argentina; 
2, LAFEX, Centro Brasileiro de Pesquisas F{\'\i}sicas, Rua Dr. Xavier Sigaud, 150, 22290-180 Rio de Janeiro, Brazil; 
3, Universidade do Estado do Rio de Janeiro, Instituto de F{\'\i}sica,
Rua S\~{a}o Francisco Xavier, 524, 20559-900 Rio de Janeiro, Brazil; 
4, Institute of High Energy Physics, P.O. Box 918, Beijing 100039, People's Republic of China; 
5, Universidad de los Andes, Dept. de Fisica, HEP Group, Apartado Aereo 4976, Bogot\'{a}, Colombia; 
6, Institute of Particle and Nuclear Physics, Center for Particle Physics, Faculty of Mathematics and Physics, Charles University in Prague, V Holesovickach 2, CZ-18000 Prague 8, Czech Republic; 
7, Institute of Physics of the Academy of Sciences of the Czech Republic, Center for Particle Physics, Na Slovance 2, CZ-18221 Prague 8, Czech Republic; 
8, Universidad San Francisco de Quito, P.O. Box 17-12-841, Quito, Ecuador; 
9, Laboratoire de Physique Subatomique et de Cosmologie,       
IN2P3-CNRS, Universite de Grenoble 1, 53 Ave. des Martyrs,
F-38026 Grenoble, France; 
10, CPPM, IN2P3-CNRS, Universit\'e de la M\'editerran\'ee, 163 Ave. de Luminy,
F-13288 Marseille, France; 
11, Laboratoire de l'Acc\'el\'erateur Lin\'eaire, IN2P3-CNRS, BP 34, 
Batiment 200, F-91898 Orsay, France; 
12, LPNHE, Universit\'es Paris VI and VII, IN2P3-CNRS,         
4 place Jussieu, Tour 33, F-75252 Paris, France; 
13, DAPNIA/Service de Physique des Particules, CEA, Saclay, 
F-91191 Gif-sur-Yvette, France; 
14, Universit{\"a}t Freiburg,  Physikalisches Institut, Hermann-Herder-Str. 3,
79104 Freiburg, Germany; 
15, Panjab University, Department of Physics, Chandigarh 160014, India; 
16, Delhi University, Department of Physics and Astrophysics, Delhi 110007, India; 
17, Tata Institute of Fundamental Research, School of Natural Sciences,
Homi Bhabha Rd., Mumbai 400005, India; 
18, University College Dublin, Department of Experimental Physics, Faculty of Science, Belfield, Dublin 4, Ireland; 
19, CINVESTAV, Departamento de F\'{\i}sica, P.O. Box 14-740, 07000 Mexico City, Mexico; 
20, FOM-Institute NIKHEF and University of                     
Amsterdam/NIKHEF, P.O. Box 41882, 1009 DB Amsterdam, The Netherlands; 
21, University of Nijmegen/NIKHEF, P.O. Box 9010, NL-6500 GL Nijmegen, The Netherlands; 
22, Joint Institute for Nuclear Research, P.O. Box 79, 
141980 Dubna, Russia; 
23, Institute for Theoretical and Experimental Physics, 
B. Cheremushkinskaya ul. 25, 117259 Moscow, Russia; 
24, Moscow State University, Department of Physics, Vorobjovy Gory,
119899 Moscow, Russia; 
25, Institute for High Energy Physics, 142284 Protvino, Russia;
26, Lancaster University, Department of Physics, Lancaster LA1 4YB, United Kingdom; 
27, Imperial College London, Department of Physics, Prince Consort Road, London SW7 2BW, United Kingdom; 
28, University of Arizona, Department of Physics, P.O. Box 210081, Tucson, Arizona 85721, USA; 
29, Lawrence Berkeley National Laboratory and University of California, 
1 Cyclotron Road, Berkeley, California 94720, USA; 
30, California State University, Department of Physics, 2345 E. San Ramon Ave., Fresno, California 93740, USA; 
31, University of California, Department of Physics and Astronomy,
4129 Frederick Reines Hall, Irvine, California 92697, USA; 
32, University of California, Department of Physics, Riverside, California 92521, USA; 
33, Florida State University, High Energy Physics-B159, Tallahassee, Florida 32306, USA; 
34, Fermi National Accelerator Laboratory, P.O. Box 500, Batavia, Illinois 60510, USA; 
35, University of Illinois at Chicago, Department of Physics,
845 W. Taylor, Chicago, Illinois 60607, USA; 
36, Northern Illinois University, Department of Physics, DeKalb, 
Illinois 60115, USA; 
37, Northwestern University, Department of Physics and Astronomy,
2145 Sheridan Road, Evanston, Illinois 60208, USA; 
38, Indiana University, Department of Physics, 727 E. 3rd St.,
Bloomington, Indiana 47405, USA; 
39, University of Notre Dame, Department of Physics, Notre Dame, Indiana 46556, USA; 
40, Iowa State University, Department of Physics, High Energy Physics Group, Ames, Iowa 50011, USA; 
41, University of Kansas, Department of Physics and Astronomy, 
1251 Wescoe Hall Dr., Lawrence, Kansas 66045, USA; 
42, Kansas State University, Department of Physics, Manhattan, Kansas 66506, USA; 
43, Louisiana Tech University, Department of Physics, Ruston, Louisiana 71272, USA; 
44, University of Maryland, Department of Physics, College Park, Maryland 20742, USA; 
45, Boston University, Department of Physics, 590 Commonwealth Ave., Boston, Massachusetts 02215, USA; 
46, Northeastern University, Department of Physics, Boston, Massachusetts 02115, USA; 
47, University of Michigan, Department of Physics, 500 E. University Ave.,
Ann Arbor, Michigan 48109, USA; 
48, Michigan State University, Department of Physics and Astronomy, East Lansing, Michigan 48824, USA; 
49, University of Nebraska, Department of Physics and Astronomy, Lincoln, Nebraska 68588, USA; 
50, Columbia University, Department of Physics, 538 W. 120th St., New York, New York 10027, USA; 
51, University of Rochester, Department of Physics and Astronomy,
Rochester, New York 14627, USA; 
52, State University of New York, Department of Physics and Astronomy, Stony Brook, New York 11794, USA; 
53, Brookhaven National Laboratory, Physics Department, Bldg. 510C, Upton, New York 11973, USA; 
54, Langston University, Department of Mathematics, Langston, Oklahoma 73050, USA; 
55, University of Oklahoma, Department of Physics and Astronomy, Norman, Oklahoma 73019, USA; 
56, Brown University, Department of Physics, 182 Hope St., Providence, Rhode Island 02912, USA; 
57, University of Texas, Department of Physics, Box 19059, Arlington, Texas 76019, USA; 
58, Rice University, Bonner Nuclear Lab, P.O. Box 1892, Houston, Texas 77005, USA; 
59, University of Virginia, Department of Physics, Charlottesville, Virginia 22901, USA; 
60, University of Washington, Department of Physics, P.O. Box 351560, Seattle, Washington 98195, USA 

\newpage

\end{document}